\documentclass[useAMS,usenatbib]{mn2e}
\usepackage[]{hyperref}
\hypersetup{colorlinks=true, linkcolor=red, citecolor=blue}
\usepackage{chngpage}
\usepackage{soul}
\usepackage{ulem}
\usepackage{color}
\usepackage{wasysym}
\usepackage{amssymb}
\usepackage{graphicx}
\usepackage{float}
\usepackage[utf8]{inputenc}

\def\la{\raise.5ex\hbox{$<$}\kern-.8em\lower 1mm\hbox{$\sim$}}
\def\ga{\raise.5ex\hbox{$>$}\kern-.8em\lower 1mm\hbox{$\sim$}}
\def\be{\begin{equation}}
\def\ee{\end{equation}}
\def\ba{\begin{eqnarray}}
\def\ea{\end{eqnarray}}

\def\Mdotstar{\dot{M}_\ast}

\def\Omegadot{\dot{\Omega}}

\def\Mdotin{\dot{M}_{\mathrm{in}}}

\def\Edot{\dot{E}}
\def\Pdot{\dot{P}}

\def\Msun{\mathrm{M}_{\astrosun}}

\def\rlc{r_{\mathrm{LC}}}
\def\rout{r_{\mathrm{out}}}
\def\rco{r_{\mathrm{CO}}}

\def\Lx{L_{\mathrm{x}}}

\def\Md{M_{\mathrm{d}}}

\def\rA{r_{\mathrm{A}}}

\def\Tp{T_{\mathrm{p}}}

\def\dM*{\delta M_*}

\def\Teff{T_{\mathrm{eff}}}

\def\P0min{P_{0,{\mathrm{min}}}}

\def\tc{\tau_{\mathrm{c}}}

\def\Alfven{Alfv$\acute{e}$n}

\def\Nsd{N_{\mathrm{SD}}}
\def\Nsu{N_{\mathrm{SU}}}

\def\Ndip{N_{\mathrm{dip}}}

\def\j{PSR J1852+0040}
\def\rx{RX 0822.0--4300}
\def\1e{1E 1207.4--5209}
\def\unitL{erg~s$^{-1}$}
\def\unitPdot{s~s$^{-1}$}
\def\Bdip{B_{\mathrm{dip}}}
\def\tsn{\tau_{\mathrm{SNR}}}
\def\Mdotcrit{\dot{M}_{\mathrm{crit}}}

\title[CCOs: some of them could be spinning up?]{Central Compact Objects: some of them could be spinning up?}

\author[Benli \& Ertan]{
O. Benli\thanks{E-mail:onurbenli@sabanciuniv.edu} and \"{U}. Ertan
\\
Sabanc\i\ University, 34956, Orhanl\i\, Tuzla, \.Istanbul, Turkey
}

\date{Accepted XXX. Received YYY; in original form ZZZ}

\pubyear{----}

\begin{document}
\label{firstpage}
\pagerange{\pageref{firstpage}--\pageref{lastpage}}
\maketitle

\begin{abstract}
Among confirmed central compact objects (CCOs), only three sources have measured period and period derivatives. We have investigated possible evolutionary paths of these three CCOs in the fallback disc model. The model can account for the individual X-ray luminosities and rotational properties of the sources consistently with their estimated supernova ages. For these sources, reasonable model curves can be obtained with dipole field strengths $\sim$ a few $\times 10^9$~G on the surface of the star. The model curves indicate that these CCOs were in the spin-up state in the early phase of evolution. The spin-down starts, while accretion is going on, at a time $t \sim 10^3 - 10^4$ yr depending on the current accretion rate, period and the magnetic dipole moment of the star. This implies that some of the CCOs with relatively long periods, weak dipole fields and high X-ray luminosities could be strong candidates to show spin-up behavior if they indeed evolve with fallback discs.  

\end{abstract}




\section{INTRODUCTION}

Central Compact Objects (CCOs) are young isolated neutron stars located close to the centers of supernova remnants. There are currently $\sim 10$ confirmed CCOs with X-ray luminosities $\Lx \sim 10^{33}$~\unitL  (see e.g. \citealt{Gotthelf_etal_13, DeLuca_17} for a review of CCOs). The periods and period derivatives, measured only for three CCOs, are in the ranges $P \sim 0.1-0.4$~s and $\Pdot \sim 10^{-17} - 10^{-18}$~\unitPdot~which give rotational powers $\Edot = I \Omega \Omegadot \sim 10^{31}-10^{32}$~\unitL, much below their $\Lx$. This indicates that their X-ray luminosities cannot be powered by rotational energy loss. Due to extremely low $\Pdot$ values the observed periods are likely to be similar to the initial periods. The surface dipole fields deduced from the dipole torque formula, $\Bdip = 3.2 \times 10^{19} \sqrt{P \Pdot}$, are around $10^{10}$~G. Their X-ray spectra can be modelled with two blackbody spectra with emitting areas much smaller than the surface area of the neutron star \citep{halpern_gotthelf_10}. The sources have not been detected in the optical, infrared (IR) and radio bands yet. There are upper limits for the IR luminosities \citep{wang_etal_07, DeLuca_etal_11}.

1E1207.4-5209 (PSR J1210-5226) was discovered as an X-ray source with the \textit{Einstein} satellite \citep{helfand_becker_84}. It is located near the centre of G296.5+10, a $\sim 7$ kyr old \citep{roger_etal_88, pavlov_etal_02} supernova remnant (SNR) at a distance $d \sim 2$~kpc \citep{giacani_etal_00}. The source slows down with $P = 428$~ms and $\Pdot = 2.2 \times 10^{-17}$ s s$^{-1}$ \citep{Gotthelf_etal_13}, which give $\Edot = 1.1 \times 10^{31}$~\unitL, and the characteristic age $\tc = P/{2 \Pdot} = 3 \times 10^8$ yr. The bolometric luminosity was estimated as $\Lx \simeq 2.5 \times 10^{33}$~erg~s$^{-1}$ for $d = 2$~kpc \citep{deluca_etal_04}. The X-ray spectrum has an absorption feature at 0.7 keV and harmonics \citep{bignami_etal_03}. If these are interpreted as electron cyclotron lines, the required surface field strength is $\sim 8 \times 10^{10}$~G.

\rx~was discovered near the center of Puppis A SNR \citep{petre_etal_96}. It has $P = 112$~ms and $\Pdot = 7.3 \times 10^{-18}$~\unitPdot~\citep{Gotthelf_etal_13}, which give $\Edot = 2 \times 10^{32}$~\unitL~and $\tc = 2.4 \times 10^8$~yr. The estimated SNR age, $\tsn$,  of \rx~is in the $\sim 3.7 - 5.2$~kyr range. From the X-ray spectral analysis \citep{Gotthelf_etal_13}, the bolometric luminosity is estimated as $\Lx = 5.0 - 5.6 \times 10^{33}$~\unitL~for $d = 2.2$~kpc (see e.g. \citealt{becker_etal_12}, \citealt{Gotthelf_etal_13} for uncertainties in distance and $\tsn$ estimations). In our analysis, we adopt this $\Lx$ range for comparison with model calculations.     

\j~ (CXOU J185238.6+004020) was discovered at the center of the Kes 79 SNR \citep{seward_etal_03}. The source is quite similar to \rx~in rotational properties with $P = 104$~ms \citep{gotthelf_etal_03} and $\Pdot \simeq 8.6 \times 10^{-18}$~\unitPdot~\citep{halpern_gotthelf_10}. These measurements give $\Edot = 3 \times 10^{32}$~\unitL~and $\tc = 1.9 \times 10^8$~yr. The bolometric luminosity is estimated to be $\Lx = 5.3 \times 10^{33}$~\unitL~from the two blackbody fits to the X-ray spectrum \citep{halpern_gotthelf_10} with $d = 7.1$~kpc \citep{frail_clifton_89}. The estimated $\tsn \sim 5.4 - 7.5$~kyr is consistent with the lower limit, $\tsn > 3.2$~kyr, estimated from the ionization timescale measurements \citep{sun_etal_04}.

The three CCOs summarised above have some common distinguishing properties: (1) The dipole moments inferred from the dipole torque formula are much smaller than those of other young neutron stars, (2) the characteristic ages estimated if these sources isolated rotating dipoles are much larger than the SNR ages, (3) the observed X-ray luminosities of these sources are found to be greater than the cooling luminosities \citep{Page_etal_06} estimated for the ages indicated by their SNR, and (4) pulsed radio emission from the CCOs has not been detected yet.

Evolution of neutron stars with fallback discs could explain the observed diversity of single neutron star populations, namely CCOs, high--magnetic field radio pulsars (HBRPs), anomalous X-ray pulsars/soft gamma repeaters (AXPs/SGRs), dim isolated neutron stars (XDINs) and rotating radio transients (RRATs), as a result of the differences in the initial conditions defined by the initial period, the magnetic dipole moment, and the disc properties \citep{Alpar_01}.

Detailed analyses of AXP/SGRs \citep{Benli_Ertan_16}, XDINs \citep{Ertan_etal_14} and HBRPs (\citealt{Benli_Ertan_17, benli_ertan_18} see also \citealt{Caliskan_etal_13}) showed that the individual source properties of these populations can be produced in the same long-term evolution model, initially developed by \cite{Ertan_etal_09}, with very similar main disc parameters. In line with these results, it was shown that the characteristic phase dependent high energy spectra of AXP/SGRs can be produced consistently with the observed pulse profiles \citep{Trumper_etal_10, Trumper_etal_13, Kylafis_etal_14}. The dipole field strengths of XDINs, AXP/SGRs and HBRPs, estimated in this model, range from  $\sim 10^{11}$~G to a few $10^{12}$~G, much smaller than those inferred from the dipole torque formula, which yields $B \sim 10^{13} - 10^{15}$~G magnetar strength surface dipole fields for these classes of sources.

The signature of a fallback disc is the characteristic emission from optical and near-infrared (NIR) to mid-infrared (MIR) wavelengths. The observed fluxes depend on the inclination of the disc with respect to the line of sight of the observer, mass-flow rate of the disc,  irradiation strength, inner and outer radii of the disc. This requires a detailed study of each source separately. In particular, we showed earlier that the broad band emission from optical to MIR observed from 4U 0142+61 can be reproduced with the emission from an active and irradiated fallback disc \citep{Ertan_etal_07a}. For some other AXPs  observed only in NIR bands, the measured fluxes are also in agreement with the model spectra \citep{Ertan_Caliskan_06}.  

For the sources with low X-ray luminosities like CCOs or XDINs, the irradiation flux and the resultant emission from the disc could be below the detection limits depending also on the inclination of the disc and the distance of the source \citep{ertan_etal_17}. We estimate that most of  the young isolated neutron star systems with low luminosities, including transient AXPs, XDINs and CCOs, could be observed at NIR and MIR bands with future \textit{James Webb Space Telescope} (JWST). The IR spectra estimated for CCOs will be studied in an independent work.

In the present work, we concentrate on the properties of the three CCOs with known $P$, $\Pdot$ and $\Lx$. We trace the initial conditions through many simulations in the fallback disc model to determine the allowed ranges of initial conditions that can reproduce the individual observed properties ($P$, $\Pdot$, $\Lx$). In Section 2, we briefly describe the model and give the results of its applications to the three CCOs. We discuss the results and summarise our conclusions in Section 3.

\section{THE MODEL AND THE APPLICATIONS TO CCOs}

Details of the long-term evolution model used in the present work was described in the earlier work with applications to AXP/SGRs \citep{Benli_Ertan_16} and XDINs \citep{Ertan_etal_14}. Here, we just describe the main disc parameters, the initial conditions, and summarise the evolutionary phases.  

There are many unknowns about the formation of fallback discs (angular momentum and mass of fallback matter, and the conditions that make the disc active or passive). Given these uncertainties, if the evolution with a fallback disc explain the properties of a particular source, we simply assume that the initial conditions were convenient for the formation of an extended disc around this source. We determine the allowed ranges of disc mass and the dipole field from the simulations.

The initial conditions of a given source are defined by the initial period ($P_0$), the initial disc mass ($\Md$), and the strength of the magnetic dipole field on the pole of the star ($B_0$). The disc diffusion equation is solved for a standard thin disc using the $\alpha$ prescription for the kinematic viscosity \citep{Shakura_Sunyaev_73}. The kinematic viscosity parameter, $\alpha$, the X-ray irradiation efficiency, $C$, and the minimum critical temperature, $\Tp$, of the active disc are the main disc parameters which are expected to be similar for the fallback discs of different neutron star populations within the simplifications of our model. It is the differences in the initial conditions creating the different evolutionary paths leading to diverse properties.

When the source is in the accretion phase, $\Lx$ is produced by mass-flow on to the star, while the intrinsic cooling of the star is the source of $\Lx$ in the propeller phase, when accretion is not allowed. The outer radius, $\rout$, of the active disc is the radius where $\Teff$ is currently equal to the critical temperature, $\Tp$, below which the disc becomes passive \citep{Balbus_Hawley_91, Inutsuka_Sano_05}. This dynamical $\rout$ decreases gradually in time with decreasing $\Lx$. It was shown earlier that the properties of XDINs, AXP/SGRs and HBRPs can be reproduced with $\alpha = 0.045$, $\Tp \sim 100$~K and $C = 1 - 7 \times 10^{-4}$ \citep{Ertan_etal_14, Benli_Ertan_16, Benli_Ertan_17}. 


\begin{figure}
\centering
\includegraphics[width=\columnwidth,angle=0]{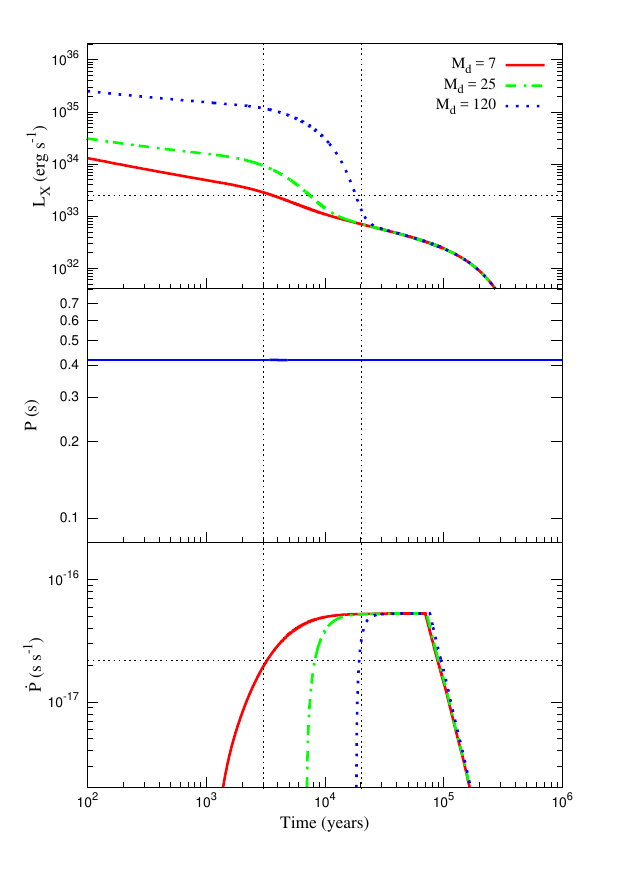}
\caption{ Illustrative model curves for \1e. From the top to the bottom, the panels shows the $\Lx$, $P$ and $\Pdot$ evolutions. In the top panel, we also present cooling curve (dotted-curve) to show the discrepancy between cooling luminosity and $\Lx$ at SNR ages.The observed source properties (horizontal dashed-lines) are reached in the estimated range of SNR age (vertical dashed-lines) with $B_0 = 4.2 \times 10^9$~G. Different model curves are obtained with different disc mass as given in units of $10^{-7}~\Msun$ in the top panel. The asymptotic growth in $\Pdot$ values until $t \sim 10^4$~yr, is due to decreasing spin-up torque with decreasing $\Mdotstar$. Beginning of sharp decrease in $\Pdot$ at $t \sim (7-8) \times 10^4$~ yr corresponds to the termination of the accretion phase and the onset of propeller phase.  }
\label{fig:1207}
\end{figure}

We calculate the magnetic spin-down torque exerted by the disc on the star by integrating the magnetic stresses from the conventional \Alfven~radius $\rA = (G M)^{-1/7}~\mu^{4/7}~\Mdotin^{-2/7}$ to the co-rotation radius, $\rco$, where $\mu$ and $M$ are the magnetic dipole moment and mass of the neutron star, $G$ is the gravitational constant and $\Mdotin$ the rate of the mass flow to the inner disc. This spin-down torque can be related to $\Mdotin$ and $\rA$ through 

\be\label{sd_torque}
\Nsd \simeq (G M \rA)^{1/2}~\Mdotin~\left[1-\left(\frac{\rA}{\rco}\right)^3\right]
\ee

\noindent \citep{Ertan_Erkut_08}. When the estimated $\rA$ is greater than $\rlc$, we replace $\rA$ with $\rlc$. In the accretion with spin-down (ASD) phase, there is also a spin-up torque, $\Nsu = (G M \rco)^{1/2} \Mdotstar$, arising from the mass-flow on to the star. Except at very short periods, contribution of the magnetic dipole torque, $\Ndip$, to spin-down is negligible in comparison with the disc torque. In the numerical calculations, we include the effect of $\Ndip$ and $\Nsu$ as well. The spin-up torque $\Nsu$ is also found to be negligible in the long-term evolution of AXP/SGRs and XDINs. For CCOs studied in this work, however, we find that $\Nsu$ dominates $\Nsd$ in the very early phase of evolution for about $10^3-10^4$ yr due to relatively weak dipole fields of CCOs. Nevertheless, we find that this early spin-up phase does not affect the subsequent evolution of the source significantly (see Section 3).


\begin{figure}
\centering
\includegraphics[width=\columnwidth,angle=0]{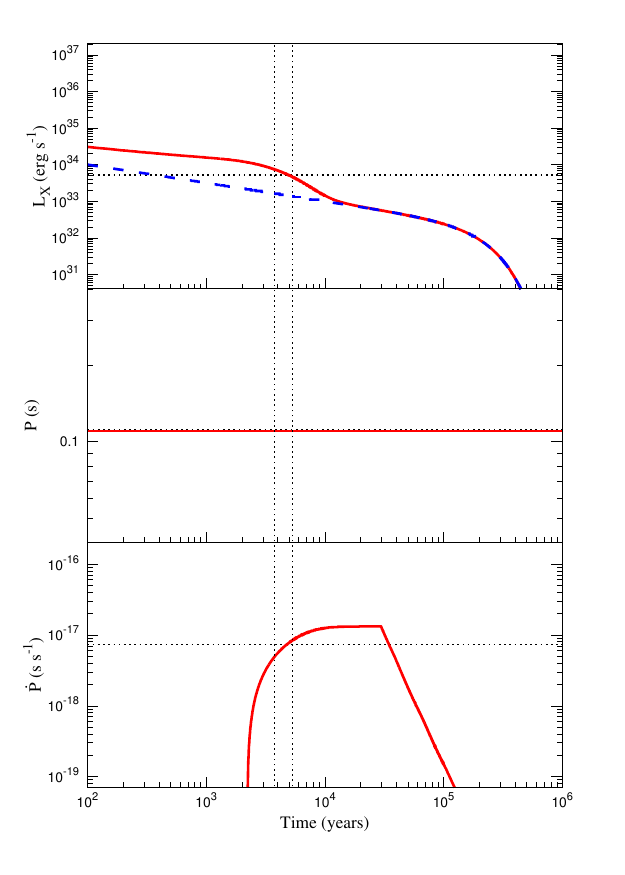}
\caption{ The model curve for the long-term evolution of \rx. This evolution is obtained with $B_0 = 2.1 \times 10^{9}$ G and $\Md = 2.5 \times 10^{-6}~\Msun$. The observed $\Lx$, $P$ and $\Pdot$ is reached at the time $t \sim 4-5 \times 10^3$~yr.    }
\label{fig:0822}
\end{figure}

\begin{figure}
\centering
\includegraphics[width=\columnwidth,angle=0]{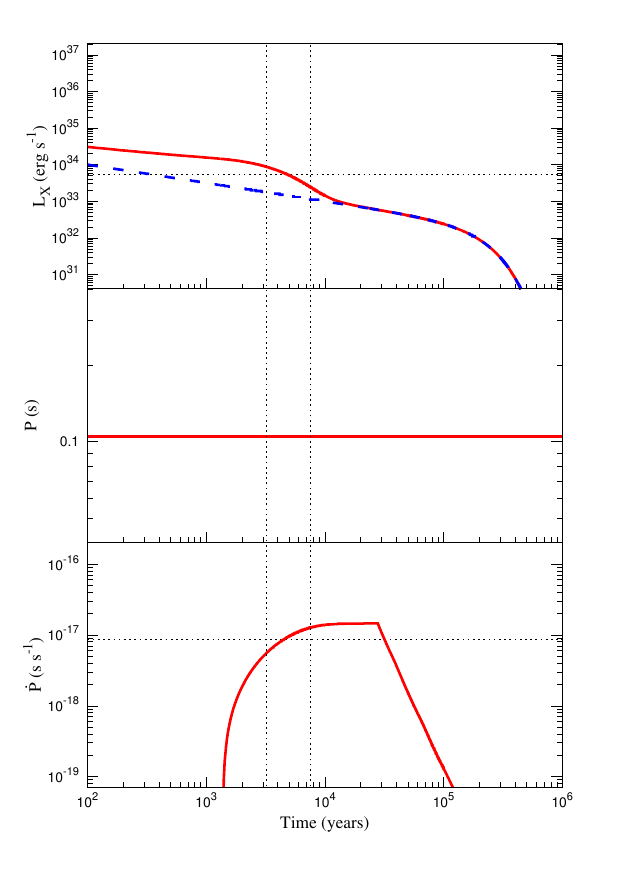}
\caption{The same as Fig. \ref{fig:0822}, but for \j. The model curve is produced with $B_0 = 2.2 \times 10^{9}$~G and $\Md = 3 \times 10^{-6}~\Msun$.} 
\label{fig:1852}
\end{figure}

In the ASD phase, we take $\Mdotstar = \Mdotin$, and assume that $\Mdotstar = 0$ in the propeller phase. For the transition from ASD phase to the strong propeller phase, we use the simplified condition $\rA > \rlc$ due to the lack of a detailed estimation for the critical condition. In our model, since this transition takes place in the late, sharp decay phase of $\Mdotin$, the uncertainty in the transition rate $\Mdotcrit$ does not affect our results significantly. Our transition condition is roughly in agreement with the critical transition rate $\Mdotcrit$ estimated recently by \cite{Ertan_17} which is consistent with the estimated luminosities of the transitional millisecond pulsars (tMSPs) during the transitions between the X-ray pulsar and the radio pulsar states (see e.g \citealt{Jaodand_etal_16}). We note that the minimum accretion rates producing the X-ray pulsations of tMSPs in the X-ray pulsar states are orders of magnitude smaller than the rate required for $\rA = \rco$ which is usually assumed to  be the condition for the accretion-propeller transition in conventional models \citep{illarionov_sunyaev_75}.

We have applied this model to the three CCOs with measured $P$, $\Pdot$ and $\Lx$, namely \j, \rx, \1e, with typical main disc parameters ($\alpha=0.045$, $\Tp = 100$~K and $C = 1 \times 10^{-4}$) used in the earlier applications of the model. The model curves given in Figs. \ref{fig:1207}-\ref{fig:1852} show that the model can reproduce the individual properties of CCOs ($P$, $\Pdot$, $\Lx$) simultaneously. These model curves imply that the sources started their evolutions in the spin-up phase, and enter the spin-down phase at later times depending on the initial conditions. The estimated ages of the sources in the model are in agreement with their $\tsn \simeq 10^3-10^4$~yr. The properties of  \1e~can be obtained with a large range of $\Md$, but with a narrow range of $B_0 \sim 4 \times 10^9$~G (Fig. \ref{fig:1207}). This field is smaller than the value inferred from the cyclotron line interpretation of observed absorption features ($\sim 8 \times 10^{10}$~G). We note that the surface field strength could differ from the pure dipole field in the presence of local quadrupole fields (see e.g. \citealt{Guver_etal_11, Tiengo_etal_13}). On the other hand, the dipole field inferred from the dipole torque formula is similar to that estimated from the absorption lines. Nevertheless, evolution of a neutron star with this dipole field in vacuum does not produce source properties consistently with estimated SNR ages. The model curves in Fig. \ref{fig:1207} illustrate evolutions for different $\Md$ values with the same $B_0 \sim 4 \times 10^9$~G. The uncertainties in the estimated SNR ages for \rx~and \j~are relatively small, which constrains their $\Md$ as well. The evolutionary curves seen in Figs. \ref{fig:0822} and \ref{fig:1852} that could represent the evolutions of \rx~and \j~are both obtained with $B_0 \sim 2 \times 10^9$~G and $\Md \sim 3 \times 10^{-6}~\Msun$. There are only small differences in their $B_0$ and $P_0$ parameters.

The model can produce the properties of these three CCOs in the accretion regime, when they are powered by mainly the accretion luminosities. Theoretical cooling curves estimated for neutron stars with conventional fields (dashed curves in Figs \ref{fig:1207}-\ref{fig:1852}) remain below observed $\Lx$ (at $t \simeq \tsn$) for all these sources.

\section{CONCLUSIONS}

We have investigated the long-term evolutions of \1e, \rx, \j, the CCOs with measured period derivatives. Estimated SNR ages of these three CCOs are not consistent with the cooling ages corresponding to their X-ray luminosities. Conventional cooling curves for neutron stars give luminosities that are several times smaller than the observed $\Lx$ of the sources at $t \approx \tsn$. We have shown that the individual source properties ($P$, $\Pdot$, $\Lx$) of these sources could be simultaneously reached by neutron stars evolving with fallback discs. We find the sources in the accretion phase at ages that are in agreement with estimated supernova ages. For the three CCOs, we have obtained reasonable model curves with the initial conditions $B_0 \sim (2-4) \times 10^9$~G and $\Md \sim 10^{-6}-10^{-5}~\Msun$. For the main disc parameters ($\alpha$, $\Tp$, $C$), we take values similar to those used in our earlier applications to AXP/SGRs, XDINs and HBRPs (see Figs. \ref{fig:1207}-\ref{fig:1852}).

In our model, unlike AXP/SGRs, XDINs and HBRPs, CCOs start their evolution in the spin-up phase which lasts $\sim 10^3-10^4$~yr. The magnetic dipole fields of CCOs estimated in the model are much weaker than those of XDINs ($B_0 \gtrsim 10^{11}$~G) and AXP/SGRs ($B_0 \gtrsim 10^{12}$~G). This seems to produce a gap between the $B_0$ distributions of CCOs  and XDINs estimated in our model. Nevertheless, our preliminary results for RRATs show that this gap (a few $10^9$~G - $10^{11}$~G) could be filled by the $B_0$ distribution of RRATs. This will be studied in an independent work.

The occurrence of $B_0 \sim 10^{9}$~G initial surface dipole fields does not seem to be very common among the young isolated neutron stars. Such weak dipole moments are typical of X-ray MSPs (neutron stars in low-mass X-ray binaries (LMXBs)) and radio millisecond pulsars \citep{Manchester_etal_05}. However as the birth rate of CCOs corresponds to only a fraction of the birth rate of young neutron stars, the majority of the MSP-LMXB population still require an evolutionary decay of the dipole magnetic moment \citep{bhattacharya_heuvel_91}. 

\newpage  
\section*{Acknowledgements}

We acknowledge research support from T\"{U}B\.{I}TAK (The Scientific and Technological Research Council of Turkey) through grant 117F144 and from Sabanc{\i} University. We thank M. A. Alpar for his useful comments. 


\bsp	
\bibliographystyle{mn2e}
\bibliography{benli}

\label{lastpage}

\end{document}